\documentstyle[12pt,psfig,axodraw,a4]{article} 
\def\bild#1#2{    
        \vspace*{-5mm}
        \begin{center}
        \begin{math}
        \epsfxsize#2cm
        \epsffile{#1}
        \end{math}
        \end{center}
        }
\newcommand{\vs}{\vspace{-0.25cm}}

\begin{document} 

\begin{center}
\large{\bf \begin{boldmath}$pp\to p p\omega$ \end{boldmath} reaction 
near threshold}

\bigskip 

\bigskip

N. Kaiser\\

\bigskip

Physik Department T39, Technische Universit\"{a}t M\"{u}nchen,\\
    D-85747 Garching, Germany

\end{center}

\bigskip

\bigskip

\begin{abstract}
We analyze the total cross section data for $pp \to pp\omega$ near threshold 
measured recently at SATURNE. Using an effective range approximation for the 
on-shell $pp$ S-wave final state interaction we extract from these data
the modulus $|\Omega| = 0.53$ fm$^4$ of the threshold transition amplitude
$\Omega$. We present a calculation of various (tree-level) meson exchange
diagrams contributing to $\Omega$. It is essential that $\omega$-emission from 
the anomalous $\omega\rho\pi$-vertex
interferes destructively with $\omega$-emission from the proton lines. The
contribution of scalar $\sigma$-meson exchange to $\Omega$ turns out to be
negligibly small. Without introducing off-shell meson-nucleon form factors the
experimental value $|\Omega|=0.53$ fm$^4$ can be reproduced with an $\omega
N$-coupling constant of $g_{\omega N}=10.7$. The results of the present 
approach agree qualitatively with the J\"ulich model. We also perform a
combined analysis of the reactions $pp\to pp\pi^0,\, pn\pi^+,\, pp\eta,\,
pp\omega$ and $pn\to pn\eta$ near threshold.   
\end{abstract}


\bigskip

\bigskip
Accepted for publication in {\it Physical Review C (brief reports)}
\vskip 1.5cm

Meson production in nucleon-nucleon collisions is considered to provide 
important information on the NN-interaction at short distances due to the
necessarily large momentum transfers involved in such reactions. There is an
on-going experimental program at the proton cooler synchrotrons IUCF
(Bloomington), CELSIUS (Uppsala) and COSY (J\"ulich) to measure in detail
various mesonic final states $(\pi,\, \pi\pi,\,K, \, \eta,\, \omega,\,\phi,\,
\eta'$).  In the energy region near threshold the theoretical interpretation of
the meson production process is expected to simplify considerably  since only
few  transitions (mainly those with S-waves in  the final state) will 
contribute. 

In a recent work \cite{bkm,pklam} we have developed a novel (and rather simple)
approach to meson production in proton-proton collisions, $pp\to pp\pi^0,\, pn 
\pi^+,\,pp \eta,\,p\Lambda K^+$, near threshold. One starts from the invariant 
T-matrix at threshold in the center-of-mass frame which is parameterized in 
terms of one (or two) constant threshold amplitudes. Close to threshold the 
relative momentum of the nucleons in the final state is very small and their 
empirically known strong S-wave interaction plays an essential role in the
description of the meson-production data (see also ref.\cite{sibir}). In fact 
is was found in refs.\cite{bkm,pklam} that in all cases the energy dependence
of the total cross section near threshold is completely and accurately
determined by the  three-body phase space and the on-shell S-wave final state
interaction. Close to threshold the final state interaction can even be treated
in effective range approximation using the well-known values of the scattering
lengths and  effective range parameters. Note that ref.\cite{bkm} gives a
(partial) derivation of such an approach to final state interaction in the
context of effective field theory (i.e. using only Feynman diagrams). Once one
accepts such a phenomenological separation of the (on-shell) final state
interaction from the full production process, one can extract from the total
cross section data an experimental value of the constant threshold amplitude
parameterizing the T-matrix. In the next step a standard (relativistic) Feynman
diagram calculation is performed for the center-of-mass T-matrix at threshold. 
As a major result it was found in refs.\cite{bkm,pklam} that already the
well-known  tree-level (pseudoscalar and vector) meson exchange diagrams lead
to predictions for the constant threshold amplitudes which agree with the
corresponding experimental values within a few percent. 

The purpose of this brief report is to present a similar analysis for the
$\omega$-meson production channel $pp\to pp\omega$. For this reaction total
cross section data near threshold have been measured recently by the SPES3
collaboration at SATURNE \cite{hibou} and further data are expected to come
soon from COSY. Theoretical calculations of $\omega$-production in
$pp$-collisions have been performed within the J\"ulich meson exchange model of
hadronic interactions in refs.\cite{hanh1,hanh2}. We will consider here the 
same  $\omega$-production mechanisms as in refs.\cite{hanh1,hanh2}. However, 
since our approach is entirely analytical it allows in a very transparent way
to study the role of various meson exchanges and their sensitivity to the
coupling constants.   

The T-matrix for omega-meson production in proton-proton collisions $p_1(\vec 
p\,) +p_2(-\vec p\,)\\ \to p +p +\omega$ at threshold in the center-of-mass 
frame reads
\begin{equation} T^{\rm cm}_{\rm th}(pp\to pp\omega) = \Omega
\, (i \, \vec\sigma_1 - i\, \vec \sigma_2+ \vec \sigma_1 \times \vec
\sigma_2)\cdot (\vec \epsilon \times \vec p\,) \,\,, \end{equation}
where $\vec \epsilon$ denotes the $\omega$-meson polarization vector and 
$\vec p$ is the proton center-of-mass momentum with $|\vec p\,|=941.6$ MeV at
threshold. $\vec \sigma_1$ and $\vec \sigma_2$ are the spin-operators of the
two protons. The (complex) threshold amplitude $\Omega$ (of dimension fm$^4$) 
belongs to the transition $^3P_1\to \,^1S_0 s_1$. Of course the $\omega$-meson
couples to a conserved current, however, for the threshold kinematics this 
feature does not imply any constraints on eq.(1). We follow now the successful approach to $\pi$- and $\eta$-production of ref.\cite{bkm} and assume
the T-matrix to be constant in the near threshold region and the energy
dependence of the total cross section to be given by three-body phase space and
(on-shell) $pp$ S-wave final state interaction. In this case the unpolarized
total cross section for $pp\to pp\omega$ including $pp$ S-wave final state
interaction  reads   
\begin{eqnarray} \sigma_{\rm tot}(Q) &=& |\Omega|^2 ~M^4\,{ \sqrt{(Q+m_\omega)
(Q+4M+m_\omega)} \over [2\pi(Q+2M+m_\omega)]^3 }  \nonumber \\
&&  \times \int_{2M}^{2M+Q} dW\,\sqrt{(W^2-4M^2)\,\lambda(W^2,m_\omega^2,
(Q+2M+m_\omega)^2)}\,  F_p(W) \,\,, \end{eqnarray}
with $Q$  the center-of-mass excess energy. $W$ is the final state di-proton 
invariant mass and $\lambda(x,y,z)=x^2+y^2+z^2-2yz-2xz-2xy$ denotes the 
conventional K\"allen function. $M=938.3$ MeV and $m_\omega =  782$ MeV
stand for the proton and $\omega$-meson mass. The correction factor $F_p(W)$ 
due to the strong $pp$ S-wave final state interaction reads in effective range
approximation  
\begin{equation}  F_p(W) = \bigg[1 + {a_p   \over 4}(a_p +r_p) (W^2-4M^2) + 
{a_p^2 r_p^2 \over 64 }(W^2-4M^2)^2 \bigg]^{-1} \,\,,\end{equation}
with the empirical scattering length $a_p = 7.81$ fm and effective range 
parameter $r_p =2.77$ fm taken from ref.\cite{compil}. Using eqs.(2,3) for the
total cross section, a best fit to the five SATURNE data points \cite{hibou}
near threshold ($Q\leq31$ MeV) gives for the modulus of the threshold amplitude
\begin{equation} |\Omega| = 0.53\, {\rm fm}^4 \,\,, \end{equation} 
with a small total $\chi^2=2.5$. The resulting fit values of
$\sigma_{\rm tot}$ are given in table\,1 and the energy dependent cross section
is shown in Fig.\,1 for excess energies $Q\leq 34$ MeV.

\bigskip
\begin{table}[hbt]
\begin{center}
\begin{tabular}{|c|ccccc|}
    \hline
    $Q$~[MeV]&3.8&9.1&14.4&19.6&30.1 \\
    \hline
    $\sigma_{\rm tot}^{\rm exp}$~[$\mu$b]&$0.32\pm 0.08$&$0.70 \pm 0.14$&$1.07
 \pm 0.25 $&$1.51\pm 0.30$&$1.77 \pm 0.55$ \\
    \hline
  $\sigma_{\rm tot}^{\rm fit}$~[$\mu$b]&0.23&0.67&1.11&1.54&2.36 \\
    \hline
  \end{tabular}
\end{center}
{\it Tab.\,1: Total cross sections for $pp\to pp\omega$. The 
data are taken from ref.\cite{hibou} and the fit is described in the text.}
\end{table}
\bild{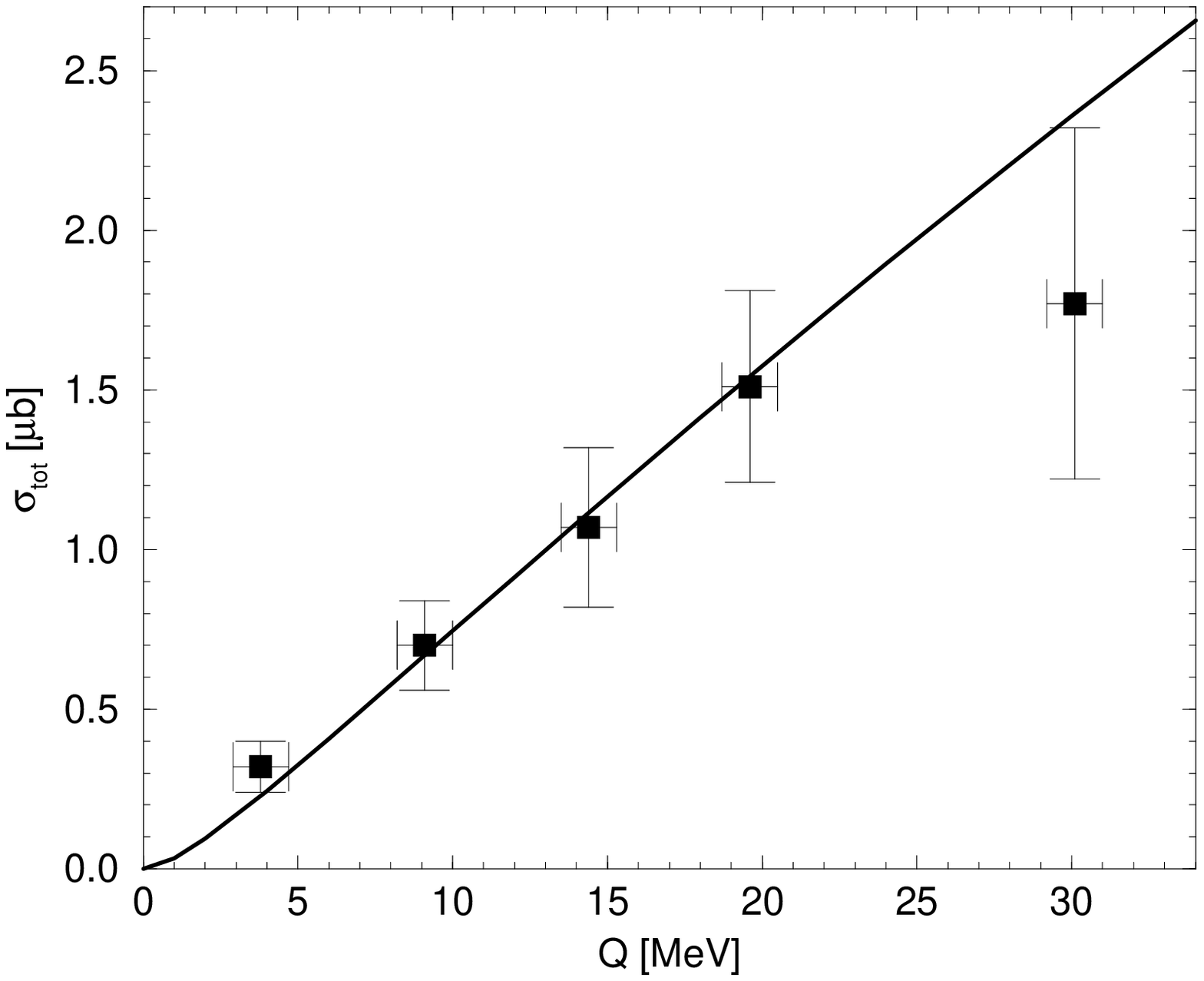}{14}
{\it Fig.\,1: Total cross sections for $pp\to pp\omega$ as a function of the
center-of-mass excess energy $Q$. The data are taken from ref.\cite{hibou} and 
the full line is calculated with $|\Omega|= 0.53$ fm$^4$ and $pp$ S-wave final
state interaction. The experimental uncertainty of $Q$ is $\pm0.9$ MeV.}  
\bigskip

\bigskip

\begin{center}
\SetScale{0.8}
\SetWidth{1.5}
  \begin{picture}(364,72)
\Line(10,0)(10,85)
\Line(80,0)(80,85)
\DashLine(10,42.5)(80,42.5){7}
\DashLine(45,42.5)(45,85){7}
\Text(20,25)[]{$\pi^0$}
\Text(50,25)[]{$\rho^0$}
\Text(45,65)[]{$\omega$}

\Line(175,0)(175,85)
\Line(246,0)(246,85)
\DashLine(175,30)(246,30){7}
\DashLine(246,50)(290,85){7}
\Text(167,33)[]{${\cal M}$}
\Text(227,55)[]{$\omega$}

\Line(350,0)(350,85)
\Line(420,0)(420,85)
\DashLine(350,55)(420,55){7}
\DashLine(420,28)(460,80){7}
\Text(309,35)[]{${\cal M}$}
\Text(372,55)[]{$\omega$}
  \end{picture}
\end{center}
\medskip
\noindent
{\it Fig.\,2: Meson exchange diagrams (${\cal M}= \pi^0, \eta, \omega, \rho^0, 
\sigma$) contributing to $pp\to pp\omega$. Graphs where the $\omega$-meson is
emitted from the other proton (solid) line and graphs with crossed outgoing
proton lines are not shown.}

\bigskip

\noindent
The quality of
reproducing the energy dependence of the data is comparable to the dynamical
calculation of ref.\cite{hanh2} where the data point at $Q=30.1$ MeV is also
somewhat overestimated. In the context of the present approach one cannot
expect the approximations eqs.(2,3) to be  valid up to $Q=30$ MeV. Note 
that in the case of $pp\to pp\pi^0$ \cite{bkm} the analogous approach based on
eqs.(2,3) has given a very good description of the total cross section data up
to $Q=21.3$ MeV. 

Next we come to the evaluation of the diagrams shown in Fig.\,2. We start with 
those graphs where the $\omega$-meson is emitted from one proton line and a 
meson is exchanged between both protons. Their contributions to $\Omega$
evidently scale with the $\omega N$-coupling constants. For the 
tensor-to-vector coupling ratio we use the value $\kappa_\omega=-0.16$ as 
found (with a small error bar) in a recent dispersion-theoretical analysis 
\cite{merg} of the nucleon electromagnetic  form factors. The for our purpose 
optimal value of the $\omega N$-coupling constant is $g_{\omega N}=10.7$. Such
a value of $g_{\omega N}$ is consistent with $g_{\omega N}=10.1 \pm 0.9$ as
obtained from forward NN-dispersion relations \cite{grein} and it is also not 
far from the one used in modern boson exchange NN-potentials \cite{jansen}.  We
find from the $\pi^0$-exchange diagrams
\begin{equation} \Omega^{(\pi)}= {g_{\omega N}g_{\pi N}^2 m_\omega  \over 
4M^2 (m_\pi^2 + Mm_\omega)(2M+m_\omega) }\bigg[1-\kappa_\omega \Big(
1+{m_\omega \over M} \Big) \bigg] = 0.42 \, {\rm fm}^4 \,\,,\end{equation}
with $g_{\pi N}=13.4$ the strong $\pi N$-coupling constant \cite{compil} and 
we used the pseudovector $\pi NN$-vertex. In the case of the pseudoscalar $\pi
NN$-vertex the expression in the square bracket would be replaced by
$1+\kappa_\omega$. Similarly, we get from  $\eta(547)$-exchange
\begin{equation} \Omega^{(\eta)}= {g_{\omega N}g_{\eta N}^2 m_\omega  \over 
4M^2 (m_\eta^2 + Mm_\omega)(2M+m_\omega) } \bigg[1-\kappa_\omega \Big(
1+{m_\omega \over M} \Big) \bigg] = 0.04 \, {\rm fm}^4 \,\,,
\end{equation}
using the SU(3)-value of the $\eta N$-coupling constant $g_{\eta N} = \sqrt3
g_{\pi N}/5=4.64$ together with the simplified ratio of the octet axial vector
coupling constants $D/F=1.5$. Since this contribution is rather small the
precise value of $g_{\eta N}$ does not matter here. Next, we find from 
$\omega$-exchange
\begin{eqnarray} \Omega^{(\omega)}&=& {g_{\omega N}^3  \over 
M m_\omega (M+m_\omega)(2M+m_\omega) }\bigg[ -\kappa_\omega + {m_\omega \over 
4M} (2+2\kappa_\omega+\kappa_\omega^2-2\kappa_\omega^3) \nonumber \\ & & -
{m_\omega^2 \over 16 M^2} \kappa_\omega^2(3+5\kappa_\omega) \bigg] = 0.28 \, 
{\rm fm}^4 \,\,, \end{eqnarray}
and from $\rho^0$-exchange
\begin{eqnarray} \Omega^{(\rho)}&=& {g_{\omega N}g_{\rho N}^2   \over 
M (m_\rho^2 + Mm_\omega)(2M+m_\omega) }\biggl[-\kappa_\omega + {m_\omega \over 
4M} (2+2\kappa_\rho+\kappa_\rho^2-2\kappa_\rho^2 \kappa_\omega) \nonumber \\ &
& + {m_\omega^2 \over 16 M^2} \kappa_\rho(\kappa_\rho-4\kappa_\omega -5
\kappa_\rho \kappa_\omega) \bigg]  = 0.74 \,  {\rm fm}^4 \,\,. \end{eqnarray}
Here, we used $\kappa_\rho=6.1$ as determined from the dispersion theoretical
analysis \cite{merg} of the nucleon electromagnetic form factors and $g_{\rho
N} =3.04$. This value follows from the $\rho$-universality relation 
$g_{\rho N} = g_{\rho \pi}/2$ with $g_{\rho \pi}=6.08$ as determined from the
empirical $\rho \to \pi\pi$ decay width. The chosen value $g_{\rho N} =3.04$
lies in between $g_{\rho N} = 2.63\pm 0.14$ as obtained from $\pi N$-dispersion
relations \cite{hopi} and $g_{\rho N} = 3.25$ as employed in the NN-potential
of ref.\cite{jansen}. Furthermore, we consider a scalar $\sigma$-meson exchange
and find  
\begin{equation} \Omega^{(\sigma)}= {g_{\omega N}g_{\sigma N}^2 \over 
M (m_\sigma^2 + Mm_\omega)(2M+m_\omega) }\bigg[\kappa_\omega +{m_\omega \over 4
M}(1+3\kappa_\omega)  \bigg] = -0.02 \, {\rm fm}^4 \,\,,\end{equation}
inserting the mass $m_\sigma=550$ MeV and the coupling constant $g_{\sigma N}=
8.5$ taken from \cite{mach}. Even though the mass $m_\sigma$ is rather
low and the coupling constant $g_{\sigma N}$ is rather large, the contribution
$\Omega^{(\sigma)}$ turns out to be negligibly small. Consequently, scalar
meson exchanges do not play any significant role in $\omega$-production near 
threshold. Obviously, one needs now a large negative contribution to cancel
the large $\rho^0$-exchange contribution $\Omega^{(\rho)}$. This can be
generated by $\omega$-emission from the anomalous $\omega\rho\pi$-vertex
\begin{equation}  {\cal L}_{\omega \rho \pi} = -{G_{\omega \rho
\pi}\over f_\pi} \, \epsilon^{\mu \nu \alpha \beta} \,( \partial_\mu \,
\omega_\nu )\, \vec \rho_\alpha \cdot \partial_\beta \vec \pi\,\,,
\end{equation} 
with $\epsilon^{\mu\nu\alpha\beta}$ the totally antisymmetric tensor in four
dimensions ($\epsilon^{0123}=-1$) and $f_\pi=92.4$ MeV the weak pion decay 
constant. Evaluation of the first diagram in Fig.\,2  leads to
\begin{equation} \Omega^{(an)} = {g_{\pi N} g_{\rho N} (1+\kappa_\rho)
G_{\omega \rho \pi} m_\omega^2 \over 4M f_\pi (m_\pi^2 +M m_\omega)(m_\rho^2 +M
m_\omega) } = -0.93 \, {\rm fm}^4 \,, \end{equation}
and we used $G_{\omega \rho \pi}=-1.2$ as determined (modulo the sign) in
\cite{klingl} from systematic studies of $\omega(782)$- and $\phi(1020)
$-decays. This value of $G_{\omega\rho\pi}$ (including the sign) is equivalent
to the one used in ref.\cite{hanh1}, where also a large cancelation between the
dominant $\omega\rho\pi$-exchange current and nucleonic current contributions
was found. Note that in our previous work \cite{bkm} on $pp\to pp\pi^0$ we took
a $G_{\omega\rho\pi}$-value of positive sign as given by the Wess-Zumino-term 
for vector meson. This specific form of ${\cal L}_{\omega \rho\pi}$ which 
assumes a particular realization of vector meson dominance (with direct
vector-meson-photon conversion) is, however, not mandatory and alternative
derivations of a Lagrangian ${\cal L}_{\omega\rho \pi}$ have been given in 
ref.\cite{jain}. In any case the contribution of the $\omega\rho\pi$-vertex to 
the $pp\to pp\pi^0$  threshold amplitude is very small (about 3\%) and a sign 
change does not matter.  Due to the large cancelation between $\Omega^{(\rho)}$
and $\Omega^{(an)}$ one observes a moderate dependence of total amplitude 
$\Omega$ on the $\rho N$-coupling constant $g_{\rho N}$. Within the range
mentioned above one finds variations of $\pm 0.05$ fm$^4$. The negative 
value of the tensor-to-vector coupling ratio $\kappa_\omega=-0.16$ is
in fact crucial, setting $\kappa_\omega=0$ the total sum of all contributions
would be reduced to $\Omega=0.30$ fm$^4$. A similar strong dependence on
$\kappa_\omega$ was observed in the J\"ulich model \cite{hanh1}. In contrast to
\cite{hanh1,hanh2} we see no need to introduce off-shell meson-nucleon form
factors in order to reproduce the empirical value of $|\Omega|=0.53$ fm$^4$.
The basic reason for this is that the final state interaction of
ref.\cite{hanh1} leads to a strong enhancement of the total cross section near
threshold,  whereas it causes a strong reduction  in the present approach
based on eqs.(2,3) since $F_p(W)\leq 1$ (see appendix C in ref.\cite{bkm}). For
a discussion of the comparably small and more uncertain effects due to nucleon
resonance excitation and other meson-exchange currents, see ref.\cite{hanh1}.

Let us finally combine the results of ref.\cite{bkm} for $pp\to pN\pi$ and $NN
\to NN\eta$ with the ones obtained here for $pp\to pp\omega$ and search for a
common set of coupling constants. For reasons of consistency we omit the small 
pion loop contribution ${\cal A}^{(loop)}=-0.14$ fm$^4$ calculated in
\cite{bkm} and we complete the expressions for ${\cal A}^{(\omega)},\,{\cal
A}^{(\omega\rho\pi)}, {\cal B}^{(\omega)},\, {\cal C}^{(\omega)}$ and ${\cal 
D}^{(\omega)}$ by the terms coming from the non-zero $\kappa_\omega$. This 
amounts to very small changes of about 1\% in the case $NN\to NN\pi$ and of 
about 5\% in the case $NN\to NN\eta$. Furthermore, we disregard the small 
$\sigma$-meson exchange contribution $\Omega^{(\sigma)}$.  With the coupling
constants $g_{\pi N}=13.4, 
\,g_{\rho N}=3.04,\, \kappa_\rho = 6.1,\, \kappa_\omega =-0.16$ taken fixed 
and $g_{\omega N}=9.8,\, g_{\eta N}=5.22,\, G_{\omega \rho\pi}=-1.03$ being 
adjusted one obtains ${\cal A}=2.74$ fm$^4$, ${\cal B}=2.69$ fm$^4$, ${\cal C}
=1.32$ fm$^4$, ${\cal D}=1.86$ fm$^4$, $\Omega =0.53$ fm$^4$ in comparison 
to the experimental values  of ${\cal A}=(2.7-0.3\,i)$ fm$^4$, ${\cal B}=(
2.8-1.5\,i)$ fm$^4$, $|{\cal C}| =1.32$  fm$^4$, $|{\cal D|}=2.3$ fm$^4$, 
$|\Omega|=0.53$ fm$^4$. With regard to the simplicity of the whole approach
this points towards a very remarkable consistency. However, such a nice 
agreement does not immediately imply that the mechanisms of meson production 
in NN-collisions are understood. It may be that the dynamical complexities of
such processes reveal themselves only in the more exclusive observables like 
angular distributions of differential cross sections and asymmetries generated 
by polarized beams and targets. We hope to report on these topics in the near 
future.

\end{document}